\documentclass[12pt,twocolumn]{emulateapj}

\usepackage{color}
\usepackage[usenames,dvipsnames]{xcolor}
\usepackage[colorlinks,urlcolor=blue,citecolor=blue,linkcolor=blue]{hyperref}
\usepackage{amsmath}
\slugcomment{}

\begin{document} 

\title{An Infrared Diffuse Circumstellar Band? \\ The Unusual 1.5272 Micron DIB In the Red Square Nebula} 
\shorttitle{Anomalous IR DIB in the Red Square Nebula}

\author{G.~Zasowski\altaffilmark{1,2}, 
S.~Drew~Chojnowski\altaffilmark{3},
D.~G.~Whelan\altaffilmark{4}, \\
A.~S.~Miroshnichenko\altaffilmark{5},
D.~A.~Garc\'{i}a~Hern\'{a}ndez\altaffilmark{6,7},
S.~R.~Majewski\altaffilmark{8}
}
\shortauthors{Zasowski et al.}

\altaffiltext{1}{NSF Astronomy and Astrophysics Postdoctoral Fellow; gail.zasowski@gmail.com}
\altaffiltext{2}{Department of Physics \& Astronomy, Johns Hopkins University, Baltimore, MD, 21218, USA}
\altaffiltext{3}{Apache Point Observatory and New Mexico State University, Sunspot, NM 88349, USA}
\altaffiltext{4}{Department of Physics, Austin College, Sherman, TX 75090, USA}
\altaffiltext{5}{Department of Physics \& Astronomy, University of North Carolina at Greensboro, Greensboro, NC 27402, USA}
\altaffiltext{6}{Instituto de Astrof\'{i}sica de Canarias, E-38205 La Laguna, Tenerife, Spain}
\altaffiltext{7}{Departamento de Astrof\'{i}sica, Universidad de La Laguna, E-38206 La Laguna, Tenerife, Spain}
\altaffiltext{8}{Department of Astronomy, University of Virginia, Charlottesville, VA, 22904, USA}

\begin{abstract}
The molecular carriers of the ubiquitous absorption features called the diffuse interstellar bands (DIBs) have eluded identification for many decades, 
in part because of the enormous parameter space spanned by the candidates and the limited
set of empirical constraints afforded by observations in the diffuse interstellar medium.
Detection of these features in circumstellar regions, where the environmental properties 
are more easily measured,
is thus a promising approach to understanding the chemical nature of the carriers themselves.
Here, using high resolution spectra from the APOGEE survey, 
we present an analysis of the unusually asymmetric 1.5272~$\mu$m DIB feature along the sightline to the Red Square Nebula 
and demonstrate the likely circumstellar origin of about half of 
the DIB absorption in this line of sight.
This interpretation is supported both by the velocities of the feature components 
and by the ratio of foreground to total reddening along the line of sight.
The Red Square Nebula sightline offers the unique opportunity to study the behavior of DIB carriers in a constrained environment
and thus to shed new light on the carriers themselves.
\end{abstract}


\section{Introduction} \label{sec:intro}

For nearly a century, absorption features known as the diffuse interstellar bands (DIBs) have been
observed superimposed on the spectra of stars, galaxies, and quasars along numerous lines of sight through the Milky Way
and other galaxies \citep[e.g.,][]{Heger_1922_stationarylines,Cox_2011_DIBssummary}.  
The current strongest consensus is that most (if not all) of the distinct features are produced by
carbon-dominated molecules, such as polycyclic aromatic hydrocarbons (PAHs) or fullerenes 
\citep[e.g.,][]{Herbig_1995_dibs,Sarre_2006_DIBssummary},
but so far, definitive association of any specific species with a specific DIB has proven elusive.

The vast majority of, perhaps all, DIB detections have occurred in the 
diffuse interstellar medium (ISM), unassociated with stars \citep[][]{Seab_1995_DCBs}.
However, circumstellar environments are particularly interesting potential hosts of DIB carriers,
because often different types of information concerning their temperature, density, and chemistry are available than for
a particular patch of diffuse interstellar cloud at a generally unknown distance.
Furthermore, as circumstellar environments are typically significantly warmer and denser than diffuse ISM clouds, they
provide additional, complementary regimes in which to characterize DIB behavior and thus their carriers.
Searches for DIB features in circumstellar environments have been 
either unsuccessful or inconclusive in a variety of sources, from protostars to planetary nebulae
\cite[PNe; e.g.,][]{LeBertre_1993_DCBs,Oudmaijer_1997_YSO-DIBs,GarciaHernandez_2013_fullerenedibs}.
Slightly more promising candidates have been found by
\citet{DiazLuis_2015_fullereneDCBs}, who noted absorption features towards the PN Tc~1
at wavelengths consistent with DIBs blue-shifted within the range of the nebular emission velocity;
\citet{Dahlstrom_2013_herschel36dibs} and \citet{Oka_2013_Herschel36DIBs}, 
interpreted some unusually broad and asymmetric DIBs towards the infrared (IR) source Herschel~36
as most likely due to excitation of the DIB carrier molecules by the local radiation field, but see also \citet{Bernstein_2015_6614DIB}.
Broadening and asymmetry may also be caused by the superposition of DIB features from multiple clouds 
along the line of sight, 
and many DIBs have been determined to have intrinsic substructure and asymmetry 
\citep[e.g., 6203\AA\, and 13175\AA;][]{Porceddu_1991_6200angDIBs,Rawlings_2014_JbandDIBs}.

Here we present the discovery of a 
strongly asymmetric instance of
the (normally symmetric) 1.5272~$\mu$m IR DIB \citep[vacuum wavelength;][]{Geballe_2011_IRdibs,Zasowski_2015_dibs}
along the line of sight towards the Red Square Nebula \citep[RSN, also MWC~922;][]{Merrill_1949_mwc}.
The RSN,
a luminous IR source variously classified
as a proto-planetary nebula and a B[e] star \citep{Lamers_1998_Be-classification,Meixner_1999_PPNcandidates},
has strong hydrogen  
and metallic emission lines that provide 
constraints on the geometrical and kinematical structure of the nebula 
\citep{Andrillat_1976_emissionline-IRexcess,Tuthill_2007_redsquare,Whelan_2014_mwc922}.
The circumstellar environment hosts a substantial amount of dust, atomic, and molecular material
\citep{Andrillat_1976_emissionline-IRexcess,vanDiedenhoven_2004_PAHprofiles},
and we demonstrate here that it is also responsible for approximately half of the 1.5272~$\mu$m DIB absorption
along this line of sight, based on evidence from foreground reddening and the absorption feature shape and velocity.

In Section~\ref{sec:data}, 
we describe the data used 
and the extraction of the RSN's 1.5272~$\mu$m DIB profile.
In Sections~\ref{sec:analysis}-\ref{sec:other_interpretations}, 
we present the properties of this unusual detection, compare it to the typical profiles observed in the ISM,
briefly examine other IR and optical DIBs along this line of sight, and explore plausible interpretations.
In Section~\ref{sec:conclusions}, 
we summarize our conclusions.

\section{Data} \label{sec:data}

\subsection{$H$-Band Spectra} \label{sec:apogee_spectra}

The primary data used are high-resolution, high signal-to-noise 
$H$-band spectra of the RSN observed by the Apache Point Observatory Galactic Evolution Experiment (APOGEE) 
survey \citep[][2015 in prep]{Majewski_2012_apogee}.
APOGEE is a high resolution ($R \sim 22,500$), $H$-band ($\lambda = 1.51-1.70~\mu$m) 
spectroscopic survey studying the chemodynamical evolution
of the Milky Way (MW) with a large sample of predominantly red giant stars 
\citep{Zasowski_2013_apogeetargeting}. 
The survey uses a custom-built 300-fiber spectrograph \citep{Wilson_2012_apogee} on the 2.5-m Sloan telescope \citep{Gunn_2006_sloantelescope};
public release of the data began in 2013 with the Sloan Digital Sky Survey III \citep[SDSS-III;][]{Eisenstein_11_sdss3overview}
Data Release 10 \citep[DR10;][]{Ahn_2014_dr10}
and culminated in 2015 with DR12 \citep{Alam_2015_SDSSDR12}.

\begin{figure}[!hptb]
\begin{center}
   \includegraphics[trim=0.1in 4.6in 3in 1.5in, clip, angle=0, width=0.5\textwidth]{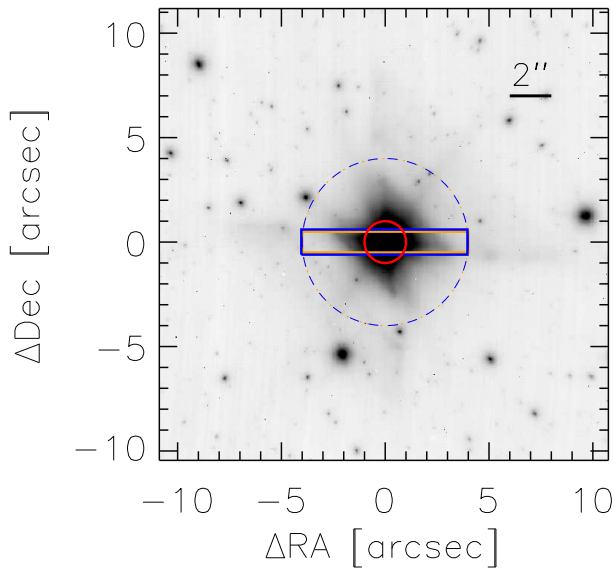} 
\end{center}
\caption{
Keck NIRC2 $H$-band image of the RSN (grayscale), 
with the apertures of the 2$^{\prime\prime}$ APOGEE fiber (red circle),
a $1.2^{\prime\prime} \times 8^{\prime\prime}$ optical slit (outer blue rectangle),
and a $1.0^{\prime\prime} \times 8^{\prime\prime}$ optical slit (inner orange rectangle).
}
\label{fig:RSN_image}
\end{figure}

Because of their usefulness in identifying terrestrial atmospheric (telluric) absorption features,
several thousand early type stars were also observed by the survey. 
The RSN was observed as part of this sample on 4, 23, and 25 May 2013.
Figure~\ref{fig:RSN_image} is a Keck NIRC2 $H$-band image 
of the RSN, along with the size and position of the APOGEE fiber field of view (red) and some slits (blue and orange)
used for supplementary information on this sightline's optical DIBs. 

All spectra taken with APOGEE are reduced using a custom reduction pipeline \citep{Nidever_2015_apogeereduction},
which produces an ``apVisit'' data file for each stellar observation.
This pipeline calculates the radial velocity (RV) for each visit, but as it is not optimized for emission line spectra,
we calculate our own RVs with a more robust treatment of the spectral features 
and use those to stack the apVisit data into a single source spectrum (with ${\rm S/N} > 460$).

\subsection{Extraction of 1.5272~$\mu$m DIB Profile} \label{sec:spectra_fit}

\begin{figure*}[!hptb]
\begin{center}
  \includegraphics[trim=0.5in 5.8in 0.9in 1.5in, clip, angle=0, width=\textwidth]{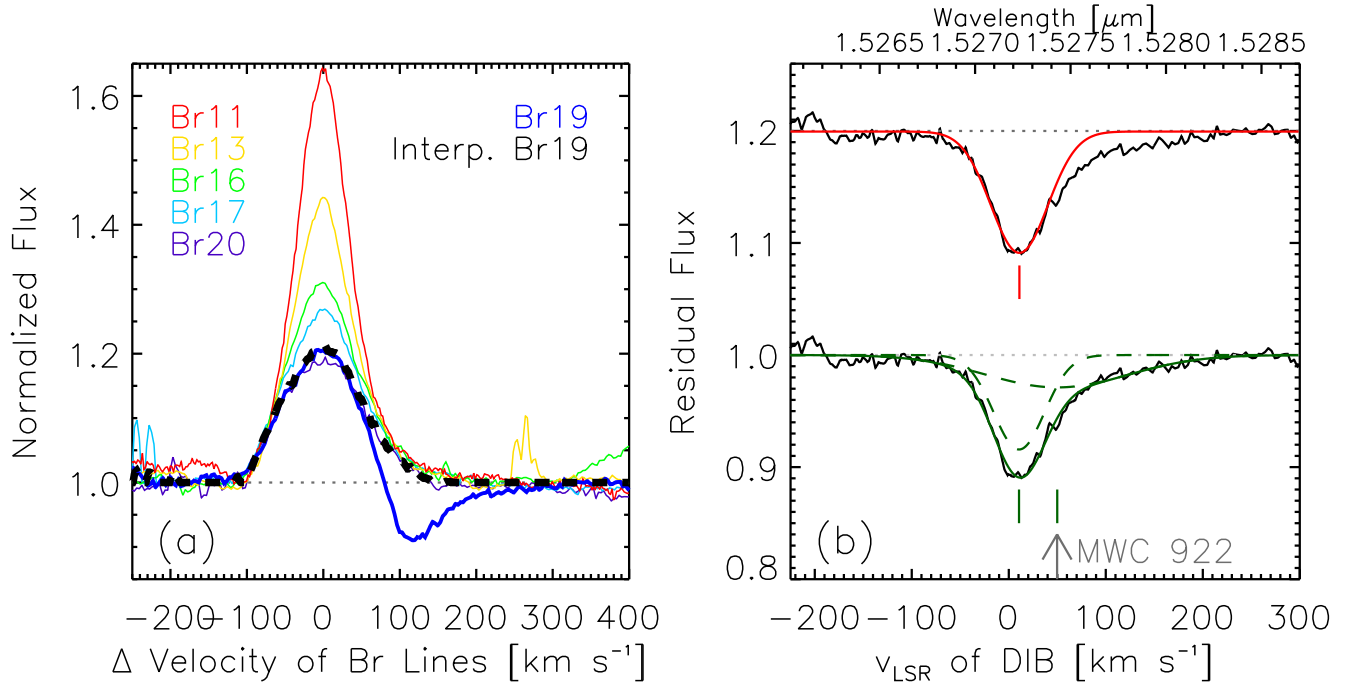} 
\end{center}
\caption{
Measurement of the overlappping Br19 and 1.5272~$\mu$m DIB profiles.
{\it (a):} The unblended Brackett profiles (thin colored lines) are used to interpolate the intrinsic Br19 profile (dashed black line).
The observed Br19+DIB profile is shown in blue.
{\it (b):} The residual DIB absorption, versus velocity $v_{\rm LSR}$, 
with single (top, red) and dual (bottom, green) Gaussian fits.
The vertical lines indicate the centers of the Gaussian components, and the gray arrow is the $v_{\rm LSR}$ of the RSN's Br19 line.
}
\label{fig:RSN_fits}
\end{figure*}

The 1.5272~$\mu$m DIB 
overlaps the Brackett-19 hydrogen emission line (Br19; $\lambda_0 = 15\,264.725$~\AA).
Because the DIB appears very close to where the Br19 line drops to the continuum, 
careful treatment of the intrinsic Br19 profile shape is required.
Here, we model the Br19 profile empirically using all of the unblended 
hydrogen lines available in the APOGEE spectrum (i.e., Br11, Br13, Br16, Br17, and Br20).
In the rest velocity frame, we calculate a quadratic relationship between the observed flux 
$F_\lambda(\Delta v)$ and the $\lambda_0$ at each $\Delta v$ of the Brackett lines.
Then, we interpolate this relationship at the Br19 rest wavelength to construct our empirical Br19 profile.
Figure~\ref{fig:RSN_fits}a shows these profiles (colored lines) and the interpolated Br19 model (black dashed line), 
along with the observed Br19+DIB profile (thick blue line).
We measure a heliocentric RV of 37.8 km~s$^{-1}$ from this line,
corresponding to a velocity relative to the Local Standard of Rest ($v_{\rm LSR}$) of 49.8~km~s$^{-1}$.

\section{Analysis} \label{sec:analysis}

Figure~\ref{fig:RSN_fits}b shows the strong residual absorption ($R_\lambda$) remaining after normalization
by this empirical Br19 profile model 
(with two fitting scenarios that are discussed in Section~\ref{sec:dib_fits}).
To our knowledge, there are no other atomic or molecular transitions at this wavelength that could
be strong and broad enough in this source to produce the absorption observed.

We calculate an integrated equivalent width $W_{\rm DIB} = 557$~m\AA\, 
by summing $(1-R_\lambda)$ in the window $-150 \le v_{\rm LSR} \le 250$~km~s$^{-1}$;
random uncertainties based on the ``flat'' span of $R_\lambda$ between $-300 \le v_{\rm LSR} \le -150$~km~s$^{-1}$ are $\lesssim$1~m\AA.
Assuming the mean ratio $W_{\rm DIB}/A_V=102$~m\AA~mag$^{-1}$ for this DIB \citep{Zasowski_2015_dibs},
and a typical $R_V=3.1$ extinction law, this equivalent width corresponds to a line-of-sight dust reddening of $E(B-V) \approx 1.75$.
At high reddening, the dispersion in the $W_{\rm DIB}/A_V$ ratio is $\lesssim$40~m\AA~mag$^{-1}$.

Of course, for any circumstellar material, the extinction law (and $W_{\rm DIB}/A_V$ ratio) may be atypical.
However, we can make some independent estimates of the interstellar reddening.
The optical DIBs also measured as part of this analysis 
suggest a total reddening of $E(B-V) \sim 1.0-1.8$~mag, with a dispersion of about 0.5~mag, 
depending on the optical spectrum used (see Section~\ref{sec:optical_dibs}).
The foreground-only reddening, as measured from PanSTARRS+2MASS photometry by 
\citet{Green_2014_PS1dustmap,Green_2015_PS1dustmap_aas}
for the $7^\prime \times 7^\prime$ region around the RSN's position, 
is $E(B-V) \approx 0.57$$^{+0.03}_{-0.08}$ and $\approx 0.87$$^{+0.05}_{-0.05}$ mag
at heliocentric distances of $\sim$2 and $\sim$3~kpc, respectively.
These lower values are consistent with the presence of {\it additional} dust along the line of sight being traced by the IR DIB,
and perhaps some of the optical ones.

\subsection{Asymmetry of Profile}

\begin{figure*}[!hptb]
\begin{center}
   \includegraphics[trim=5.5in 1.2in 1in 4.4in, clip, angle=180, width=0.45\textwidth]{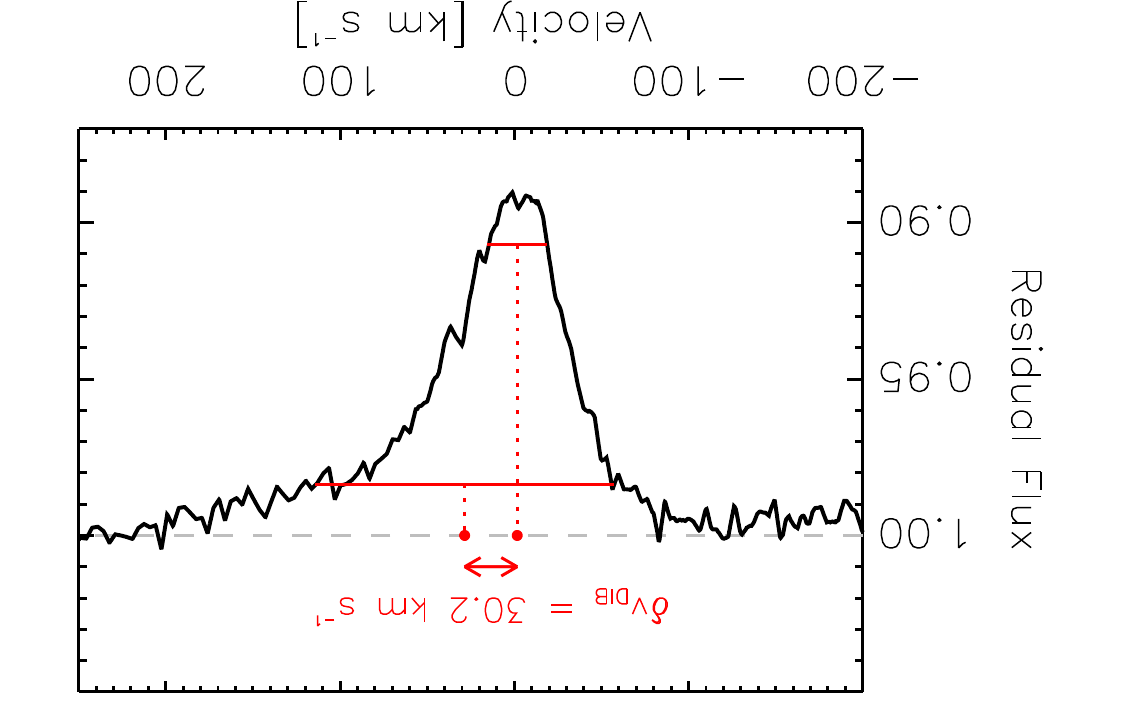} 
   \includegraphics[trim=0.75in 4.5in 5.7in 1.1in, clip, angle=180, width=0.45\textwidth]{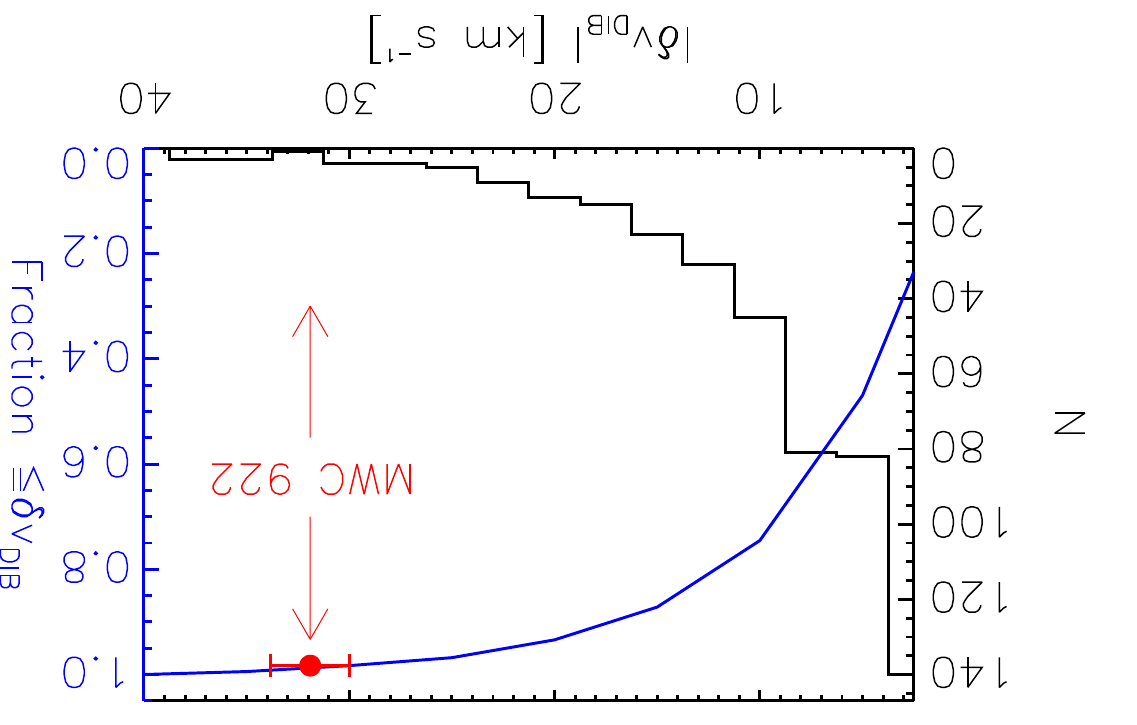} 
\end{center}
\caption{
{\it Left:} The measurement of the asymmetry estimate $\delta v$ for the RSN's DIB.
{\it Right:} The distribution of ISM $\delta v$ values, compared to that from the RSN sightline.
}
\label{fig:rsn_bis}
\end{figure*}

To quantitatively compare the asymmetry in the RSN's 1.5272~$\mu$m DIB 
to that of the typical 1.5272~$\mu$m DIB features observed in the ISM, 
we use a simplified version of the line bisector estimator
\cite[e.g.,][]{Toner_1988_starlinemeasurement},
which compares the midpoints of a profile near its top and bottom (15\% and 85\% of the amplitude; left panel of Figure~\ref{fig:rsn_bis}).
A difference in these midpoints (here, $\delta v$) indicates asymmetry in the profile.
When compared using this metric to a set of several thousand 1.5272~$\mu$m DIB features
that sample the ISM towards cool giant stars\footnote{This comparison
set comprises all sightlines from \citet{Zasowski_2015_dibs} with $\sigma(R_\lambda) \le 0.03$, 
$N_{\rm visits} = 1$ or $RV_{\rm scatter} \le 0.5$~km~s$^{-1}$, 
$\sigma(R_\lambda)/\sigma(F_\lambda) < 0.55$, and $N_{\rm pix} \ge 1350$; see Section~2.2 of that paper for details.}
at identical spectral resolution,
the RSN sightline appears more asymmetric than all but $\sim$2\% of the larger sample (right panel of Figure~\ref{fig:rsn_bis}).
Note that this comparison set does contain noisy sources, because restricting it to only those sightlines with ``good''
DIB fits would explicitly remove any asymmetric features.  Visual inspection of those $\sim$2\% asymmetric sources
confirms that the vast majority are contaminated by residual stellar features, 
though we cannot rule out a few potentially intrinsically
lopsided DIBs that will be good candidates for follow-up observations.  
None of the sightlines with potentially asymmetric 1.5272~$\mu$m DIB features lie closer than 2.4$^\circ$ to the RSN,
and none of the DIBs in the $\sim$60 sightlines in the dataset at closer projected distances 
show any signs of asymmetry.

\subsection{DIB Fit Components} \label{sec:dib_fits}

We turn now to different options for fitting this feature.
The upper $R_\lambda$ profile in Figure~\ref{fig:RSN_fits}b is overplotted in red with the 
Gaussian that best reproduces the blue (lower velocity) side of the feature; 
the extended red excess is obvious.  (Fitting the entire feature with a single Gaussian yields significantly noisier residuals.)
The bulk of the feature is well-represented by this Gaussian with a FWHM of 3.5~\AA.  
If this DIB feature has a single origin, the red excess may be due to changes in the DIB carrier's electronic properties (Section~\ref{sec:other_interpretations}).
The fact that the deepest part of the feature appears at {$v_{\rm LSR} \sim 12$~km~s$^{-1}$, 
nearly 37~km~s$^{-1}$ offset from the nebular Brackett emission, is a strike against its origin in the nebula.
The mismatch does not completely rule out a circumstellar origin, as a
range of velocities (of tens of km~s$^{-1}$) is observed in other 
absorption and emission lines definitively associated with the nebula, 
arising from 
the superposition of multiple nebular components
\citep[e.g.,][]{Whelan_2014_mwc922}.
However, a simpler solution may be sought.

The preferred scenario is shown with the lower $R_\lambda$ profile of Figure~\ref{fig:RSN_fits}b, 
overplotted in green with the best simultaneous fit dual-Gaussian components, 
which together reproduce the entire feature very well.
Each component corresponds to an approximately equal $W_{\rm DIB}$ contribution of 277~m\AA, 
and thus an $E(B-V) \approx 0.87$~mag.  This is consistent with the {\it foreground} reddening predicted by the
PanSTARRS+2MASS map in the distance range of the RSN, implying an additional reddening of 0.87 mag due to {\it circumstellar} material.
Using instead a dense environment extinction curve of $R_V = 5$ with one component, 
this additional reddening is closer to $E(B-V) \sim 0.54$~mag, 
which brings the total reddening to 1.4~mag.
Both of these are within the (albeit broad) range of reddening spanned by the optical DIBs,
suggesting that some of the optical DIB absorption may arise from the circumstellar medium as well.

The use of reddening arguments {\it alone} as evidence for circumstellar DIBs is notoriously tricky,
because of the large uncertainties and intrinsic dispersions in the DIB/$E(B-V)$ relations \citep[e.g.,][]{Luna_2008_DIBs-in-pAGBs}.  Thus,
further critical evidence for this scenario lies in the line-of-sight velocities of the two components.  
The narrower feature (${\rm FWHM} = 3.1$~\AA) has $v_{\rm LSR} = 10.3$~km~s$^{-1}$, 
well within the range spanned by the optical DIBs and the interstellar Na~D1 and K~I lines. 
In contrast, the wider feature has $v_{\rm LSR} = 49.6$~km~s$^{-1}$, 
nearly identical to that measured for the Br emission lines of the RSN itself.
Under the interpretation of a circumstellar origin, the larger width (${\rm FWHM} = 8.7$~\AA)
may be seen as simply velocity broadening due to turbulence or rotation within the nebula.
If the foreground component approximately represents the intrinsic feature width,
this implies $\sim$47~km~s$^{-1}$ of broadening within the nebula, consistent with the range of velocities
observed in different nebular components \citep[e.g.,][]{Whelan_2014_mwc922}.
Thus, if the two postulated components indeed represent different physical origins, 
then they serve as strong evidence for DIB carriers in the {\it circumstellar} environment of the RSN.

\subsection{Other $H$-Band DIBs} \label{sec:hband_dibs}

Ten known DIBs lie within APOGEE's IR wavelength coverage \citep{Geballe_2011_IRdibs,Cox_2014_xshooterDIBs},
including the 1.5272~$\mu$m feature.
Based on the small number of sightlines in which the others have been observed, 
their strengths are expected to be at most $\sim$30\% of the 1.5272~$\mu$m DIB, if detected at all.  
Despite the strong 1.5272~$\mu$m absorption in the RSN, however, 
the difficulty of measuring these weaker bands is greatly exacerbated 
(indeed, dominated) by the complex spectrum of the RSN --- 
the numerous strong, blended, and sometimes unidentified emission
lines make determination of the underlying continuum extraordinarily tricky.  
(The 1.5272~$\mu$m DIB lies in an exceptionally ``smooth'' region.)
Figure~\ref{fig:other_dibs} demonstrates this with six of the DIBs.
These features are indeed quite weak, but we do not see strong evidence for any feature asymmetries.

\begin{figure*}[!hptb]
\begin{center}
   \includegraphics[trim=0.7in 3in 2in 1.8in, clip, width=0.45\textwidth]{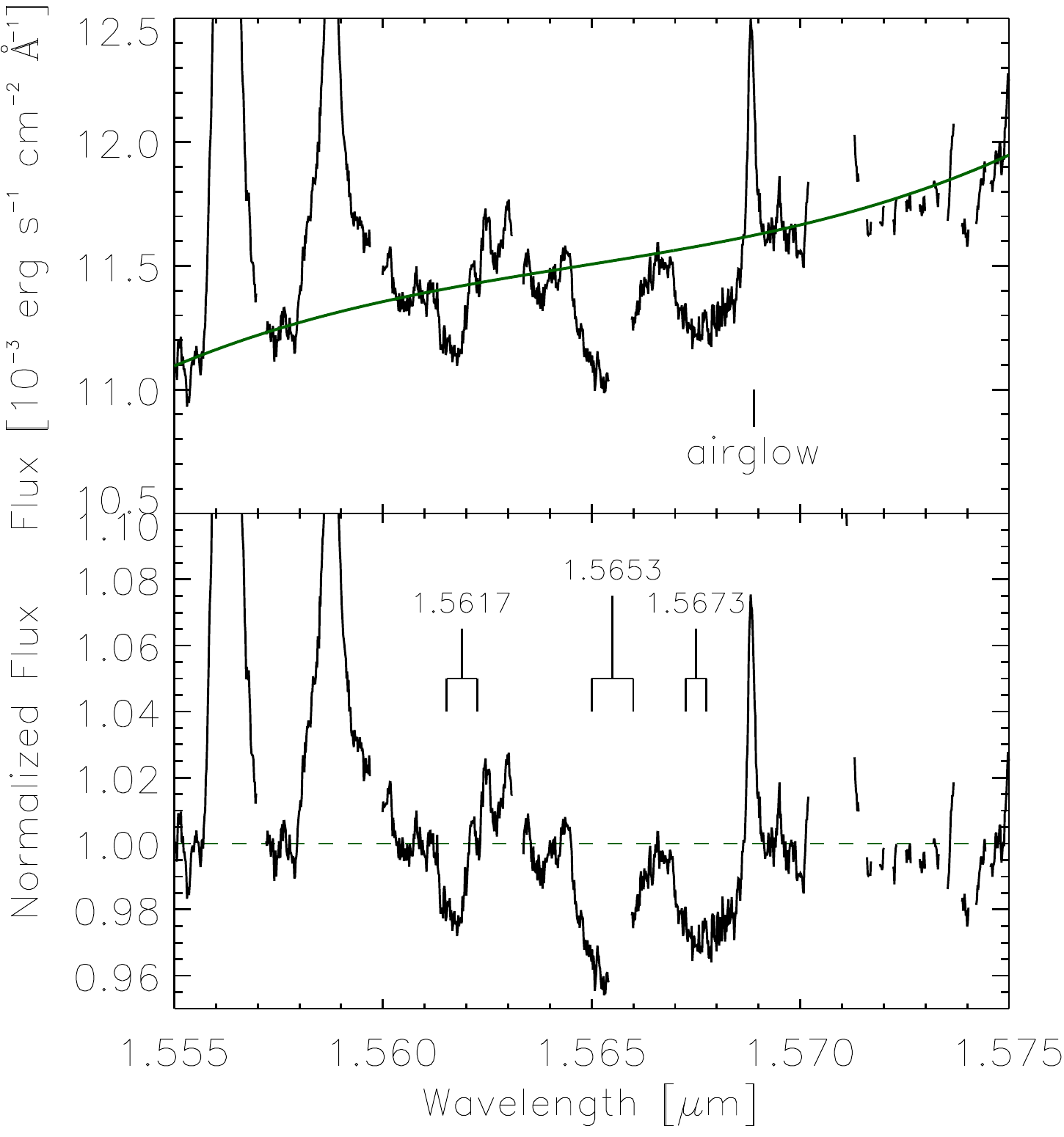} 
   \includegraphics[trim=0.7in 3in 2in 1.8in, clip, width=0.45\textwidth]{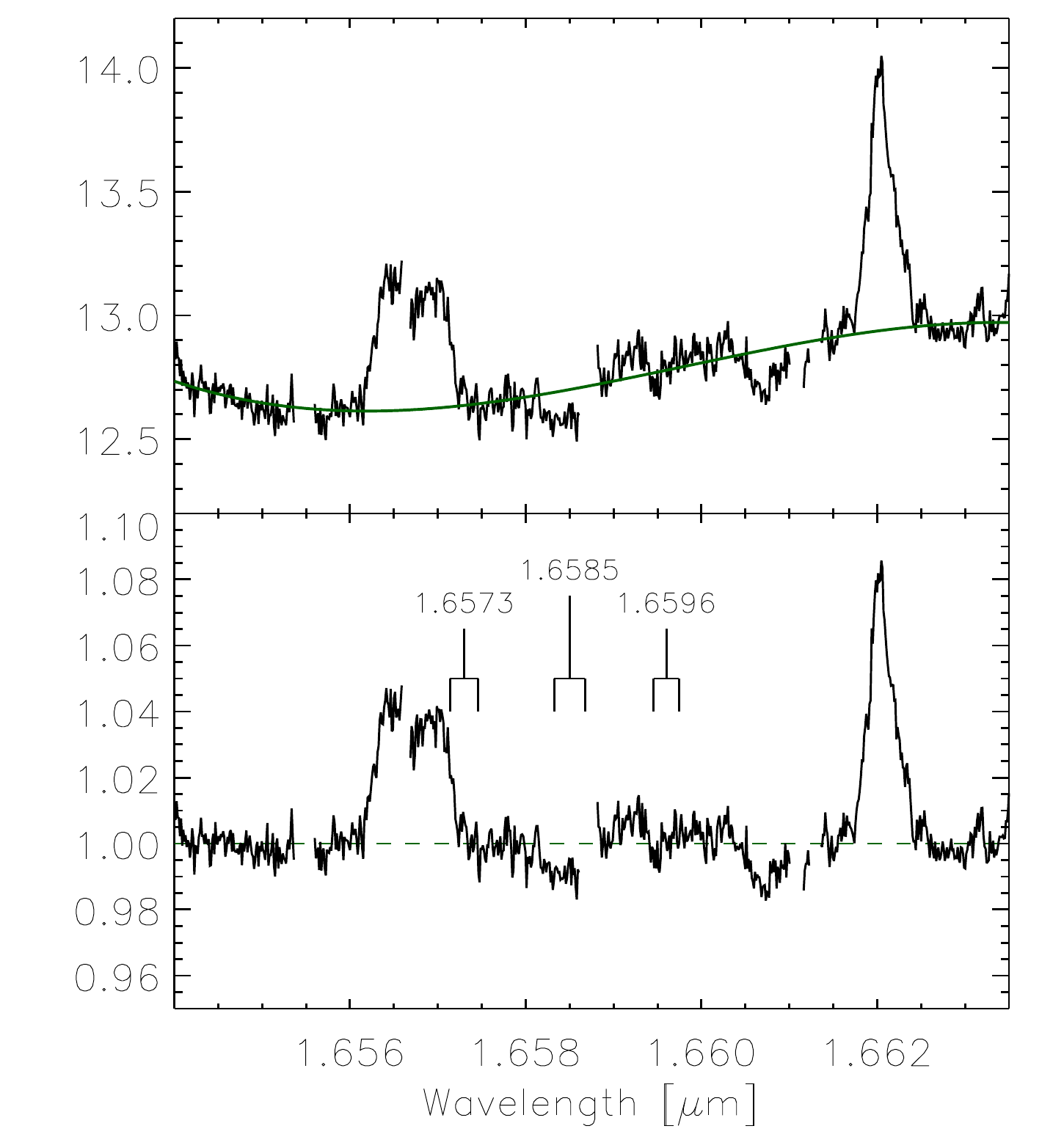} 
\end{center}
\caption{
Examples of other DIBs in APOGEE's wavelength range.  
{\it Top:} Observed flux (black) and polynomial continuum (green).
{\it Bottom:} Continuum normalized flux, with DIB central wavelengths and FWHMs
\citep[black brackets; average FWHMs from][]{Geballe_2011_IRdibs,Cox_2014_xshooterDIBs}.  
Pixels flagged as BAD or with high errors in any of the apVisit flux arrays have been masked.
}
\label{fig:other_dibs}
\end{figure*}

\subsection{Optical DIBs} \label{sec:optical_dibs}

We also studied two high-resolution optical spectra, to assess the properties of optical DIBs and interstellar lines
in the context of our proposed scenario.
One spectrum (hereafter, ``MDO'') was observed with the Harlan~J.~Smith 2.7~m telescope at McDonald Observatory 
on 6/7 September 2009, using the 
$1.2^{\prime\prime}$ slit shown in blue in Figure~\ref{fig:RSN_image}.  
The $R=60,000$ spectrum, which spans 4787--8861~\AA\, and
has been smoothed with a moving boxcar filter (with a kernel of 11 points),
has a S/N of $\sim$20 near 4860~\AA\, and $\sim$90 near 6563~\AA.
The other (hereafter, ``UVES'') was observed with the ESO VLT+UVES on 27 August 2009, using the
$1^{\prime\prime}$ slit shown in orange in Figure~\ref{fig:RSN_image}.
The UVES spectrum spans 5655--9464\AA\, with $R=42,000$ and a median S/N of 47.2.
Of the DIBs that are in regions relatively free of contaminating lines, 
we could reliably fit seven features using a simple Gaussian, shown in Figure~\ref{fig:optical_dibs} (MDO on the left, UVES on the right).

In comparing the two spectra, the largest differences are in the NaD doublet, which is saturated in the UVES
spectrum but not in the MDO one, and in the strength of the 5780\AA\, and 5797\AA\, DIB features (bottom curves in
Figure~\ref{fig:optical_dibs}).  The UVES 5780\AA\, and 5797\AA\, DIB equivalent widths are approximately twice as high
as their counterparts in the MDO spectrum, and the other DIB equivalent widths are between 1.3--1.9$\times$ as high.
The slit widths used differ only slightly (1.2$^{\prime\prime}$ for MDO and 1.0$^{\prime\prime}$ for UVES), 
and the slits were centered on the same position to within the uncertainty of the telescope pointing reported in the file headers.
To assess the possible impact of small angular offsets sampling small-scale variations in the ISM across the nebula,
we also visually examined the NaD and K~I lines from three optical VLT+X-Shooter spectra\footnote{These data, spanning 534--1020\AA, 
were obtained from the ESO archive.  One spectrum were observed on 02 July 2010 with median ${\rm S/N}=123$, and the 
other two on 07 July 2012, with median ${\rm S/N}=145$ and 144.} separated from each other by $1-2^{\prime\prime}$ on the sky.
The spectral resolution is lower, but we see no difference in the shape or strength of these interstellar features.
More detailed mapping of any ISM structure or DIB variability across the face of the nebula is reserved for future work;
here, we take the consistency of the X-Shooter spectra as evidence that the discrepancies between the MDO and UVES spectra 
are not due to strong variations in the ISM density on the scale of arcseconds.
For the remainder of this analysis, we will quote values derived from the two spectra separately.

Qualitatively, the spectra are very similar.
The critical point is that none of the DIB features, in either optical spectrum, 
display the strong asymmetry seen in the 1.5272~$\mu$m DIB
\citep[except possibly 6203\AA, which has been previously noted as intrinsically asymmetric;][]{Porceddu_1991_6200angDIBs}.
We attempted to fit two simultaneous Gaussians to these features, 
emulating the dual components seen in the 1.5272~$\mu$m DIB,
but the resulting fits were {\it quantitatively} no better than those with a single Gaussian (including for 6203\AA).
We note that at these wavelengths, the shift expected for a 37.8~km~s$^{-1}$ offset is $\sim$0.7\AA,
which is discernible even in these broad (and potentially substructured) features; 
however, there are degeneracies among the fit parameters (e.g., amplitude ratio vs.\, velocity broadening), 
and thus we cannot conclusively rule out the presence of red excess absorptions
with strength ratios or velocity properties slightly different from those of the IR DIB components.

\begin{figure*}[!hptb]
\begin{center}
   \includegraphics[trim=1in 4.9in 1.1in 1.8in, clip, width=\textwidth]{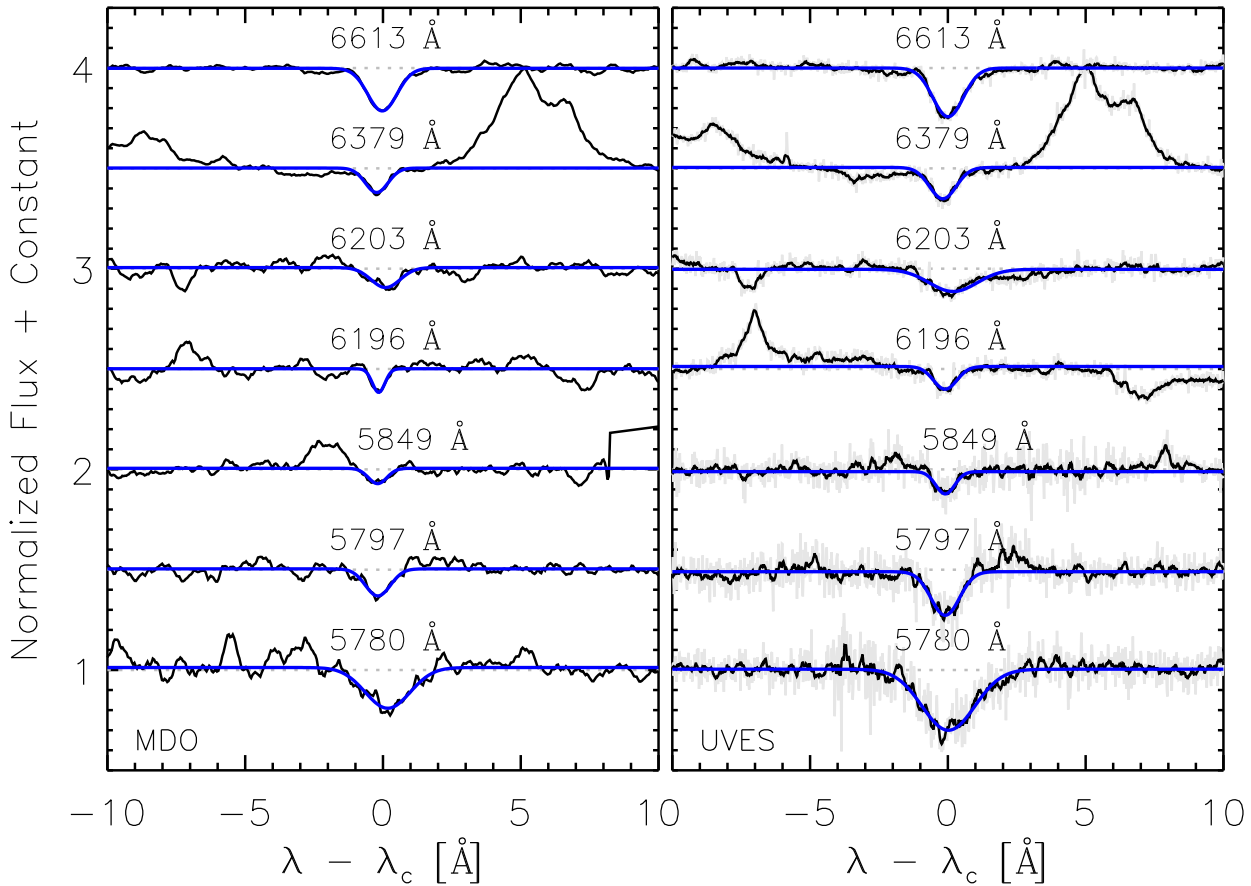} 
\end{center}
\caption{
Single Gaussian fits (blue lines) to seven optical DIBs.  The normalized spectra are offset by increments of 0.5 for visual clarity.
{\it Left:} Smoothed $R=60,000$ spectrum from McDonald Observatory.
{\it Right:} Original (gray) and smoothed (black) $R=42,000$ spectrum from VLT+UVES.
}
\label{fig:optical_dibs}
\end{figure*}

The total equivalent widths of the optical features ($W_{\rm opt}$), combined with the 
$W_{\rm opt}/W_{1.5272\mu m}$ relations calculated by \citet{Cox_2014_xshooterDIBs},
predict an absorption of the 1.5272~$\mu$m DIB that is $\sim$$30-50\pm10$\% what is observed
(from the MDO and UVES data, respectively); 
we note that these relations are based on a small number of sightlines, 
which probe interstellar material with potentially different chemistry 
or radiation exposure than that towards or in the RSN.

Using the mean $W_{\rm opt}/E(B-V)$ relations from \citet{Jenniskens_1994_dibs}, we calculate
a reddening of $E(B-V) \sim 1.0\pm0.2$~mag (MDO) and $E(B-V) \sim 1.8\pm0.5$~mag (UVES).
These are larger than the foreground reddening from the PanSTARRS+2MASS map at 2--3~kpc,
approaching that of the total estimated from the 1.5272~$\mu$m DIB.
This is consistent with a circumstellar contribution to the DIB absorption, on the order
of 15--50\% of the total.
This result on its own is merely suggestive, and we emphasize the large dispersion in the $W_{\rm opt}/E(B-V)$ relations
(both intrinsic scatter and that due to measurement uncertainties), as noted in Section~\ref{sec:dib_fits}.
However, when combined with the velocity comparisons below, 
it is worth noting the consistency of the reddening with excess DIB absorption as measured in several DIBs.

Adopting the central wavelength $\lambda_c$ values of \citet{Hobbs_2008_DIBcatalog}, 
we find DIB $v_{\rm LSR}$ values spanning $2-20$~km~s$^{-1}$, 
centered on the value measured for the ``foreground'' 1.5272~$\mu$m DIB ($v_{\rm LSR} = 10.3$~km~s$^{-1}$),
the interstellar 5895~\AA\, Na~D1 line in the MDO spectrum ($v_{\rm LSR} = 8.5$~km~s$^{-1}$),
and the interstellar 7701~\AA\, K~I line in the UVES spectrum ($<$$v_{\rm LSR}$$> = 15.2$~km~s$^{-1}$).
The optical DIB feature widths are also consistent with the widths from the interstellar cloud studies of 
\citet{Jenniskens_1994_dibs} and \citet{Hobbs_2008_DIBcatalog}, with some scatter of $\lesssim$8\%;
as noted above, though a dual-component fit does not produce a quantitatively better match to these features,
we cannot rule out the possibility that a small fraction of circumstellar material may be contributing to the observed widths.

\section{Other Interpretations} \label{sec:other_interpretations}

There exist other potential explanations for the ``additional'' long-wavelength 1.5272~$\mu$m DIB absorption,
but all face serious problems.
One interpretation is that the ISM's velocity distribution towards the RSN has a peculiar structure,
such that the ``high'' $v_{\rm LSR}$ feature is due to an unassociated ISM cloud.
In a two-cloud scenario, this would require the second one to have nearly precisely the same radial velocity as the RSN
and an internal velocity dispersion of $\gtrsim$47~km~s$^{-1}$, 
neither of which is readily apparent in the observed profiles of the optical DIBs.

To assess this in greater detail, we examined the velocity structure of the 7701~\AA\, K~I line from the UVES spectrum (Section~\ref{sec:optical_dibs}).
This is shown in the top panel of Figure~\ref{fig:ISM_velocity_model}, along with the best three-Gaussian fit
representing the three resolved subpeaks.
In the bottom panel, we ``model'' the 1.5272~$\mu$m DIB with these three components,
fixing the central velocities, velocity dispersions, and relative amplitudes, but scaling the
shortest-wavelength amplitude peak to match the deepest DIB absorption.
These components (and their summed total) are shown in red in the bottom panel of Figure~\ref{fig:ISM_velocity_model},
compared with the dual-component best fit presented in Figure~\ref{fig:RSN_fits}b and Section~\ref{sec:dib_fits}.
It is clear that while the K~I profile does reflect ISM components at higher line-of-sight velocities, 
these are not sufficient to explain the long-wavelength 1.5272~$\mu$m DIB absorption,
under the assumption that the DIB carrier and K~I are mixed in the ISM \citep[like carriers of other DIBs are; e.g,][]{Galazutdinov_2005_ISMlines}.

\begin{figure}[!hptb]
\begin{center}
   \includegraphics[trim=1.0in 5.0in 4.1in 1.8in, clip, width=0.5\textwidth]{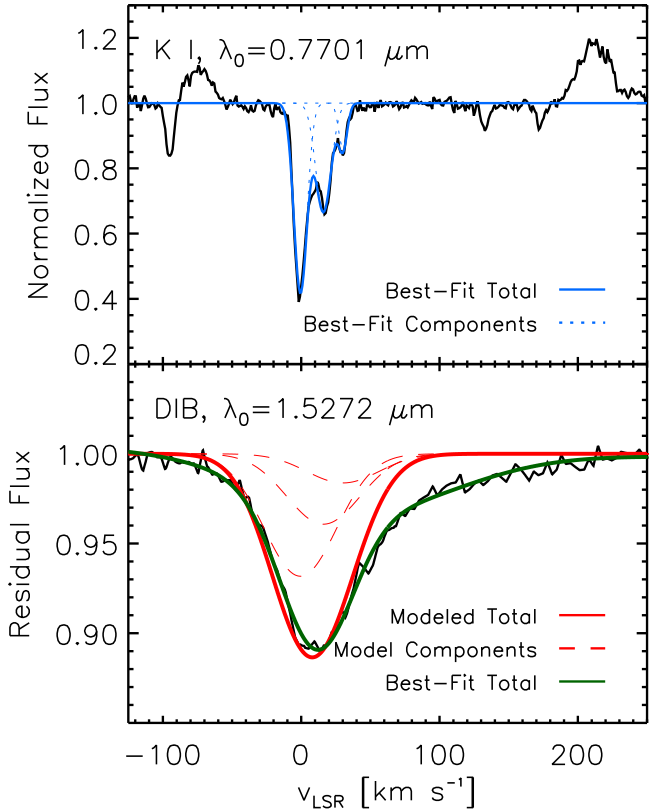} 
\end{center}
\caption{
Extrapolating a multi-component 1.5272~$\mu$m DIB from the K~I ISM line.
{\it Top:} The UVES spectrum towards the RSN, showing the clearly structured K~I line at 7701\AA.
The blue lines represent the best-fitting Gaussian components (dotted) and summed total (solid).
{\it Bottom:} The observed 1.5272~$\mu$m DIB (black), 
with the best-fit components presented in Figure~\ref{fig:RSN_fits}b and Section~\ref{sec:dib_fits} (green). 
The red dashed lines are simulated DIB components with the same central velocities, velocity dispersions,
and relative amplitudes as the K~I components above.  The solid red line is their sum.
}
\label{fig:ISM_velocity_model}
\end{figure}

Another possible explanation is the presence of an additional, 
hitherto undetected DIB blended with the 1.5272~$\mu$m feature 
\citep[e.g., as postulated by][for the Her~36 asymmetric DIBs]{Bernstein_2015_6614DIB},
or intrinsic asymmetry in this feature.
However, this is not supported by any other observation of this feature to date, 
including in heavily reddened, early-type stars with strong 1.5272~$\mu$m features that are symmetric
at a high level of confidence.
Some examples are shown in the right-hand panels of Figure~\ref{fig:comps_interp} (\ref{fig:comps_interp}b and \ref{fig:comps_interp}d).
These spectra are also APOGEE data and have a resolution identical to that shown in Figure~\ref{fig:RSN_fits}.
The red lines are single Gaussian fits.
If the long wavelength extension of the RSN's 1.5272~$\mu$m DIB were a result of an intrinsic asymmetry or a distinct DIB feature,
it should be apparent in these spectra as well, but all sightlines show a clear symmetry.

We confirmed that the asymmetry is not an artifact intrinsic to the APOGEE data by comparing the Br19+DIB profile
to that in an X-Shooter spectrum\footnote{Obtained from the ESO archive, spanning 994--2479\AA, 
observed on 02 July 2010 with a median ${\rm S/N}=538$.}; we find that they share the same shape.
Figure~\ref{fig:comps_interp} in full presents two additional examples of extracting 1.5272~$\mu$m DIB features
from the APOGEE spectra of stars with Br19 emission, using the methodology described in Section~\ref{sec:spectra_fit}.
In the same style as Figure~\ref{fig:RSN_fits}a, 
Figure~\ref{fig:comps_interp}a shows the weak, double-peaked Br emission profiles of ABE-022 \citep[][]{Chojnowski_2015_BeStars},
with the DIB clearly visible.  Figure~\ref{fig:comps_interp}c shows the same for ABE-137, a star with single-peaked emission
profiles more like the RSN.  The right-hand panel of each row contains the residual absorption after removal of the interpolated
Br19 profile as black lines, along with a single Gaussian fit (red lines).
For comparison, we also reproduce the RSN's 1.5272~$\mu$m DIB in Figure~\ref{fig:comps_interp}b in blue, and 
in Figure~\ref{fig:comps_interp}d, we overplot in green the normalized spectrum of ABE-Q09 \citep[also][]{Chojnowski_2015_BeStars},
a very early type star with no emission lines to complicate the DIB measurement.
The isolated DIB absorptions are highly symmetric, with shapes that are nearly perfectly Gaussian and that are extremely similar to the 
DIB profile in the emission-free spectrum.
This demonstrates clearly that the asymmetry in the RSN's 1.5272~$\mu$m DIB is not an artifact of our deblending process.

\begin{figure} 
\begin{center}
   \includegraphics[trim=0.5in 2.9in 0.3in 1.4in, clip, width=0.5\textwidth]{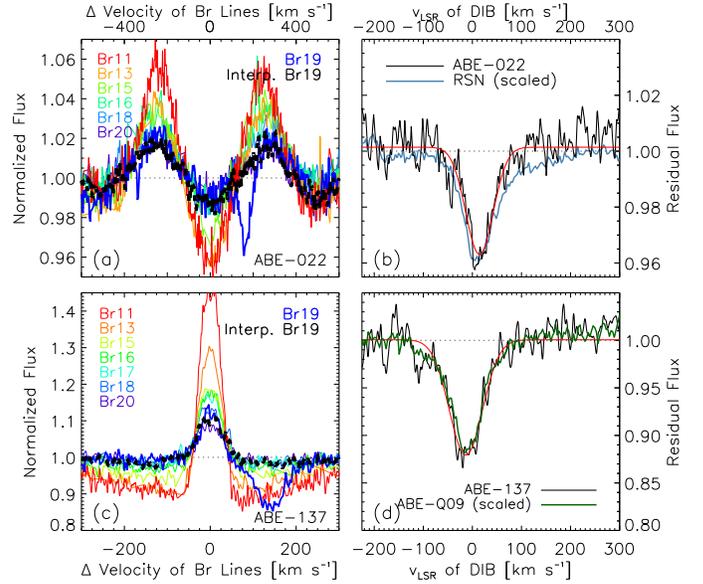} 
\end{center}
\caption{
Extraction of the 1.5272~$\mu$m DIB feature from two additional emission-line stars \citep[ABE-022 and ABE-137;][]{Chojnowski_2015_BeStars}.
Panels~({\it a}) and ({\it c}) are identical in form to Figure~\ref{fig:RSN_fits}a.  
Panels~({\it b}) and ({\it d}), like Figure~\ref{fig:RSN_fits}b, show the residual absorption in these spectra after the Br19 feature has been removed (black lines),
along with a single Gaussian fit (red lines). 
In panel~({\it b}), we also show the RSN's DIB profile (blue line) for comparison.  In panel~({\it d}),
we scale and overplot the spectrum of the early type star ABE-Q09 \citep[green line][]{Chojnowski_2015_BeStars}.
See text for details.
}
\label{fig:comps_interp}
\end{figure}

We cannot fully rule out the possibility that the RSN's 1.5272~$\mu$m DIB feature arises entirely from circumstellar carriers strongly impacted
by their local environment.
\citet{Dahlstrom_2013_herschel36dibs} reported the discovery of optical DIBs with pronounced red wings along sightlines towards Herschel~36, 
a system of O stars illuminating the M8 HII region and in close proximity to the Her~36~SE strong IR source.
Qualitatively, these wings are very similar to the one observed here in the RSN (see, e.g., their Figure 6).
Chemical models calculated by \citet{Oka_2013_Herschel36DIBs} demonstrate that these wings can be produced
when a strong radiation field drives some electrons in the DIB carrier molecules into excited states.
However, the need to invoke an outflow of nearly exactly the same velocity as the RSN's motion relative to the Sun,
such that the main feature is centered at the same velocity as the foreground ISM, 
renders this a less likely interpretation.

\section{Summary and Conclusions} \label{sec:conclusions}

We have presented the detection of a peculiar, highly asymmetric instance of the
1.5272~$\mu$m DIB along the line of sight towards the Red Square Nebula,
using high resolution $H$-band spectra from the APOGEE survey.
Strongly asymmetric profiles have never yet been observed in this particular DIB feature, or any at longer wavelengths.

The likeliest explanation is that the feature towards the RSN is a blend of absorption arising from the foreground ISM
and from the nebular material itself.  This interpretation is supported by the radial velocities of the two blended components, 
which align nearly perfectly with the foreground ISM (e.g., as traced by interstellar Na or K) 
and with the RSN's hydrogen emission lines, respectively.
In addition, the amount of interstellar reddening ``predicted'' by the interstellar DIB component nearly exactly matches
the {\it foreground} reddening traced by high-quality three-dimensional dust maps.
The {\it total} reddening predicted by the whole IR DIB feature (interstellar plus circumstellar) is 
roughly comparable than the reddening predicted by the optical DIBs, 
suggesting that at least some of the optical DIB carriers reside in the nebula itself
or that the relationship between the 1.5272~$\mu$m DIB and the optical DIBs, currently only coarsely constrained,
is substantially different in the circumstellar material.

The asymmetric 1.5272~$\mu$m absorption feature observed towards the Red Square Nebula
is likely the most confident detection to date of a diffuse circumstellar band 
\citep[DCB; see also][]{DiazLuis_2015_fullereneDCBs},
and if genuine, is also the first detection of an IR DCB,
cleanly separated from the ISM along the same line of sight in both velocity and dust reddening.
Because the circumstellar environment (its radiative properties, temperature profile, geometry, etc.) can be understood and disentangled,
this source will become a powerful laboratory for constraining the response of the DIB carriers to these environmental factors,
and thus constraining the nature of the carriers themselves.

\begin{acknowledgments}

We thank the referee for helpful comments that improved the clarity and strength of the paper.

GZ is supported by an NSF Astronomy \& Astrophysics Postdoctoral Fellowship under Award No.\ AST-1203017. 

Funding for SDSS-III has been provided by the Alfred P. Sloan Foundation, the Participating Institutions, the National Science Foundation, and the U.S. Department of Energy Office of Science. The SDSS-III web site is \url{http://www.sdss3.org/}.

SDSS-III is managed by the Astrophysical Research Consortium for the Participating Institutions of the SDSS-III Collaboration including the University of Arizona, the Brazilian Participation Group, Brookhaven National Laboratory, University of Cambridge, Carnegie Mellon University, University of Florida, the French Participation Group, the German Participation Group, Harvard University, the Instituto de Astrofisica de Canarias, the Michigan State/Notre Dame/JINA Participation Group, Johns Hopkins University, Lawrence Berkeley National Laboratory, Max Planck Institute for Astrophysics, Max Planck Institute for Extraterrestrial Physics, New Mexico State University, New York University, The Ohio State University, Pennsylvania State University, University of Portsmouth, Princeton University, the Spanish Participation Group, University of Tokyo, University of Utah, Vanderbilt University, University of Virginia, University of Washington, and Yale University. 

This paper includes data taken at The McDonald Observatory of The University of Texas at Austin, and
data obtained from the ESO Science Archive Facility under programme IDs 083.C-0676(A), 385.C-0720(A), and 089.C-0874(A). 

\end{acknowledgments}

\bibliographystyle{aa}
\bibliography{/Users/GailZasowski/Documents/APOGEE/analysis/reflib}

\end{document}